\documentclass[format=sigconf, review=false, screen=true, table, dvipsnames, preprint=true]{acmart}

\usepackage[utf8]{inputenc}
\usepackage[T1]{fontenc}

\usepackage{csquotes}
\usepackage[english]{babel}

\usepackage[flushleft]{threeparttable}
\usepackage{tabu, tabularx, booktabs}
\newcolumntype{Y}{>{\arraybackslash}X}
\newcolumntype{Z}{>{\columncolor{Gray!25}\arraybackslash}X}
\newcommand{\bftab}[1]{\fontseries{b}\selectfont#1\fontseries{\seriesdefault}}

\graphicspath{{figures/}}

\usepackage{subcaption}
\usepackage{cleveref}
\crefname{figure}{figure}{figures}
\crefname{equation}{eq.}{eqs.}

\usepackage{amsmath}
\usepackage{amssymb}
\usepackage{mathtools}
\mathtoolsset{centercolon}
\usepackage{mathbbol}

\usepackage{xspace}

\newcommand{\todo}[1]{}
\renewcommand{\todo}[1]{{\color{red} TODO: {#1}}}

\newcommand{\superscript}[1]{\ensuremath{^{\textrm{#1}}}}

\def\wc{\superscript{\dag}}
\def\bc{\superscript{\ddag}}

\newcommand{\noprf}{No-PRF\xspace}
\newcommand{\prfs}{PRF.news\xspace}
\newcommand{\prfv}{PRVF\xspace}
\newcommand{\cprfv}{C_{PRVF}\xspace}

\newcommand{\cretrieval}[2]{C_R(#1, #2)}
\newcommand{\cselection}[1]{C_{SEL}(#1)}

\newcommand{\selected}[1]{sel(#1)}
\newcommand{\numselected}[1]{{\vert \selected{#1} \vert}}

\let\emptyset\varnothing

\newcommand{\QuerySet}{Q}

\newcommand{\query}{{\MakeTextLowercase{\QuerySet}}}
\newcommand{\origQuery}{\query^\emptyset}
\newcommand{\finalQuery}{\query}
\newcommand{\expQuery}{\query^{e}}

\newcommand{\qorig}{\origQuery}
\newcommand{\qexp}{\expQuery}
\newcommand{\qfinal}{\finalQuery}

\newcommand{\corpus}{D}

\newcommand{\VerticalSet}{V}

\newcommand{\IndexVerticalSet}{{\corpus_\VerticalSet}}
\newcommand{\indexvertical}{{\corpus_\VerticalSet}}
\newcommand{\indexverticalseqlength}{{\vert \IndexVerticalSet \vert}}
\newcommand{\indexverticalseq}{\indexvertical_1 \indexvertical_2 \cdots \indexvertical_\indexverticalseqlength}

\newcommand{\set}[1]{\{#1\}}

\copyrightyear{2018} 
\acmYear{2018} 
\setcopyright{acmcopyright}
\acmConference[ICTIR'18]{2018 ACM SIGIR International Conference on the Theory of Information Retrieval}{September 14--17, 2018}{Tianjin, China}
\acmBooktitle{ICTIR '18: 2018 ACM SIGIR International Conference on the Theory of Information Retrieval, Sept. 14--17, 2018, Tianjin, China}
\acmPrice{15.00}
\acmDOI{10.1145/3234944.3234960}
\acmISBN{978-1-4503-5656-5/18/09}

\begin{document}

\title{A Vertical PRF Architecture for Microblog Search \textsuperscript{*}}
\thanks{* Please cite the ICTIR 2018 version of this paper.}

\author{Flávio Martins}
\affiliation{
    \institution{NOVA LINCS}
    \institution{Faculty of Science and Technology}
    \institution{Universidade NOVA de Lisboa}
    \postcode{2829-516}
    \city{Caparica}
    \country{Portugal}
}
\email{flaviomartins@acm.org}
\author{João Magalhães}
\affiliation{
    \institution{NOVA LINCS}
    \institution{Faculty of Science and Technology}
    \institution{Universidade NOVA de Lisboa}
    \postcode{2829-516}
    \city{Caparica}
    \country{Portugal}
}
\email{jm.magalhaes@fct.unl.pt}
\author{Jamie Callan}
\affiliation{
    \institution{Language Technologies Institute}
    \institution{School of Computer Science}
    \institution{Carnegie Mellon University}
    \streetaddress{5000 Forbes Avenue}
    \city{Pittsburgh} 
    \state{PA} 
    \postcode{15213}
    \country{USA}
}
\email{callan@cs.cmu.edu}

\renewcommand{\shortauthors}{F. Martins et al.}

\begin{abstract}
In microblog retrieval, query expansion can be essential to obtain good search results due to the short size of queries and posts.
Since information in microblogs is highly dynamic, an up-to-date index coupled with pseudo-relevance feedback (PRF) with an external corpus has a higher chance of retrieving more relevant documents and improving ranking.
In this paper, we focus on the research question: \textit{how can we reduce the query expansion computational cost while maintaining the same retrieval precision as standard PRF?}
Therefore, we propose to accelerate the query expansion step of pseudo-relevance feedback.
The hypothesis is that using an expansion corpus organized into verticals for expanding the query, will lead to a more efficient query expansion process and improved retrieval effectiveness.
Thus, the proposed query expansion method uses a distributed search architecture and resource selection algorithms to provide an efficient query expansion process.
Experiments on the TREC Microblog datasets show that the proposed approach can match or outperform standard PRF in MAP and NDCG@30, with a computational cost that is three orders of magnitude lower.
\end{abstract}

\maketitle

\section{Introduction}
\label{sec:introduction}

In microblogs there is a high mismatch between the keywords users employ to specify the information need and the words in relevant documents, which is known as the \emph{vocabulary mismatch problem}.
Query expansion is often used to increase recall, however it can also be used to produce better document rankings and increase retrieval effectiveness.
Therefore, query expansion methods based on PRF have been widely used to improve search in different collections.
It was shown to be an essential feature for microblog search \cite{lin_overview_2013}.

In standard PRF the top-$k$ documents returned by the initial query (feedback documents) are assumed to be relevant, which avoids the need for users' relevance feedback. Term weights can be calculated using collection statistics, such as in the model-based relevance models (RM) approach~\cite{lavrenko_relevance_2001}.
Several automatic query expansion methods leveraged on external static data such as dictionaries, domain-specific thesauri or precomputed corpus-specific information.
In microblogs, query expansion should be based on information that has a good coverage of real-world events.

In order to compute the expansion terms for a query using PRF, it is necessary to issue an extra initial retrieval over the whole collection, which significantly increases the computational complexity of a query.
In most production retrieval systems, caching of search results and caching of posting lists are extensively used to alleviate efficiency concerns since it significantly reduces the workload of back-end servers, especially for popular queries and query terms, and provides shorter average response times \cite{baeza-yates_efficiency_2009}.
However, even in static collections, i) query expansion is done on queries, not query terms; and ii) 20\% of all unique queries have not been seen "\textit{before}"\footnote{\url{https://europe.googleblog.com/2010/02/this-stuff-is-tough.html}}, thus, only a few queries can be cached~\cite{white_examining_2007}. This problem may be worse in dynamic collections.
Therefore, the ephemeral nature of information seeking in microblog search, calls for an architecture that provides fresh expansion terms for better retrieval results.

This paper focuses on the research question \emph{how can we reduce the query expansion computational cost while maintaining the same retrieval precision as standard PRF?}
The proposed solution brings two major advantages over standard query expansion approaches in microblogs.
First, we reformulate the query expansion process as a federated query expansion task~\cite{shokouhi_federated_2011}, and leverage on resource selection algorithms to select \emph{query-specific} information verticals.
Following the \emph{federated search}~\cite{demeester_overview_2013} terminology, a \emph{resource} can be a \emph{source} or a \emph{vertical}.
We define a \emph{vertical} (e.g., politics, technology, sports) to be a \emph{query-likelihood} search engine running on a corpus formed by the union of samples from that vertical's corresponding \emph{sources} (e.g., politico, wired, espn).
The novel vertical feedback design choice is crucial to unlock the efficiency potential of the proposed query expansion architecture.
Second, we depart from previous work that mainly use one single static information corpus and move towards an architecture where \textit{multiple information streams are constantly feeding the query expansion corpus}. 
This rationale is a step change in the way query expansion is approached: the expansion corpus is constantly updated with fresh data, it is external to the main index, and is segmented into predefined broad topics of interest.
To our knowledge, there is no prior work approaching \textit{query expansion with an external and dynamic vertical corpora}.
This paradigm shift allows a significant reduction in the document expansion data that is now limited to a few verticals of news sources and therefore allows a faster query expansion process.

The Pseudo-Relevant Vertical Feedback (PRVF),
discussed in section 3,
reduces the work-load of the whole search engine for query expansion because it selects a few news verticals and only those are then searched to retrieve feedback documents for query expansion.
Coupled with effective resource selection algorithms it
outperformed standard PRF in our experiments in section 4 and 5.

\section{Related work}

Most microblog search queries could be classified as informational: users issue a query because they are looking for more information about a subject but cannot easily clarify their query intent.
Queries submitted to Twitter were found to be significantly shorter than queries submitted to Web search engines (1.64 words vs. 3.08 words) \cite{teevan_twittersearch_2011}, therefore, query expansion is essential to provide a richer description of the information need.
The use of named entities and \emph{hashtags} in queries inspired methods that learn a feedback language model for entities~\cite{fan_improving_2015} and hashtags~\cite{efron_hashtag_2010}.
Additionally, there have been a few pseudo-relevance feedback methods proposed for microblog search that exploit temporal evidences \cite{massoudi_incorporating_2011, choi_temporal_2012, whiting_temporal_2012}.

Pseudo-relevance feedback (PRF) is an automatic query expansion technique, which was shown to significantly improve results in microblog retrieval~\cite{lin_overview_2013}.
It is a popular technique that has been applied in TREC evaluation campaigns on diverse corpora with great effectiveness.
However, it is not yet a standard feature in most production search engines, which require faster response times, because it raises efficiency issues at query time.

The standard implementation of PRF involve a two-retrieval process 1) an initial retrieval using the original query to get feedback documents and generate the expanded query, and 2) a re-retrieval using the final expanded query.
A recently proposed method, Condensed List Relevance Models (CLRM)~\cite{diaz_condensed_2015}, foregoes this expensive re-retrieval step by replacing it with the re-ranking of the original feedback documents, which can produce near identical effectiveness in traditional TREC corpora where documents are usually longer.
Previous research improves the efficiency of the whole PRF process using a precomputed document similarity matrix \cite{lavrenko_real-time_2006, cartright_fast_2010}.
However, these techniques might not be feasible in dynamic collections because term associations would need to be updated constantly.

Several studies have shown enhanced retrieval performance when leveraging on an external corpus.
\citet{diaz_improving_2006} estimate relevance models using an auxiliary large external corpus instead of the target collection and have shown improved retrieval effectiveness on several datasets.
In some cases, to improve the effectiveness of query expansion, a large external corpus that is reliable and possibly from less noisy data sources is used.
For instance, \citet{arguello_document_2008} used Wikipedia, on a blog retrieval system, and showed significant overall improvements in effectiveness over using the target blog collection for feedback.
\citet{xu_query_2009} expanded on this and proposed an approach that categorizes queries based on reliable information from Wikipedia to perform a query-dependent query expansion process based on the category detected.

\citet{bendersky_effective_2012} argue that employing multiple information sources in a number of information retrieval processes is desirable to enhanced retrieval, including query expansion.
Previous research explored the effectiveness of query expansion in federated search~\cite{shokouhi_effective_2009, ogilvie_effectiveness_2001}.
First, \citet{ogilvie_effectiveness_2001} analyzed the effectiveness of query expansion in a \enquote{one query fits all} fashion, the top-ranked documents retrieved from an index with a representative sample of all the collections are used as feedback to extract terms and create an expanded query that is then submitted to each of the selected search engines.
Later, \citet{shokouhi_effective_2009} proposed a method that uses a \emph{focused} expanded query for each selected \emph{source} (query-specialization).
These provided evidence of the feasibility of query expansion in a federated search environment effectively.
We depart from previous work to develop a scalable query expansion index architecture for PRF.
In this paper the query expansion index is external to the search collection and is partitioned into verticals to improve efficiency and effectiveness.
The architecture proposed can handle a large number of sources or topics, which can be expanded to provide a broader subject coverage.

\section{Federated Query Expansion}

We propose a novel solution that provides a balance between effectiveness and efficiency arising from a) the organization of the expansion corpus into verticals, and b) the best performing resource selection algorithms.
There are two main types of resource selection algorithms: 1) sample-document methods and 2) vocabulary-based methods. 
In sample-document methods, a \emph{centralized sample index} (CSI) is built with a representation set for each source or vertical.
Representation sets are usually a small sample of documents of about 1-10\%, which makes the resource selection process more efficient due to the small size of the resulting CSI index.
Two of the most successful sample-document methods Rank-S~\cite{kulkarni_shard_2012} and \emph{Central-rank-based collection selection} (CRCS)~\cite{shokouhi_central-rank-based_2007} use a similar strategy to the earlier ReDDE~\cite{si_relevant_2003} algorithm.
The user's query is run against the CSI and the top-$n$ retrieved documents are used in the algorithm in a voting fashion.
Vocabulary-based resource selection algorithms, such as Taily~\cite{aly_taily_2013}, represent each collection by its vocabulary statistics only.
Previous studies have shown that these approaches are highly effective in reducing the number of search engines that need to be searched.

In PRF the computational cost of the query expansion process is tied to the cost of the initial retrieval.
Tackling this major challenge, requires efficient query expansion architectures that reduce this retrieval cost and can still deliver high-quality query expansions.

\subsection{The Computational Cost of PRF}
The standard PRF procedure creates an expanded query with new terms extracted from the top-$k$ documents in the initial ranking obtained by retrieving with the original query terms.
Firstly, the user's query $\qorig$ is issued to the system to retrieve a ranked list of documents $ R(\qorig,D)$, over the whole collection $D$ using a \emph{query-likelihood} (QL) retrieval model. 
Secondly, the top-$k$ documents retrieved, $R_k(\qorig,D)$, are used to build an expansion language model $\qexp$ with the terms extracted from those documents.
Finally, the final ranking is obtained by issuing the final query $\qfinal$, which is a linear model combination of the original query language model $\qorig$ and the expansion language model $\qexp$ with parameter $\lambda$, as follows:
\begin{equation}
\qfinal = (1 - \lambda) \qorig + \lambda \qexp.
\end{equation}

The PRF cost can be expressed as the sum of two components: $C_{QE}(\qorig,D)$, the cost of retrieving the ranked list of pseudo-relevant documents using the original query $\qorig$, and $C_R(\qfinal,D)$, the cost of retrieving the final documents using the final expanded query $\qfinal$. 
Formally, the PRF cost is defined as
\begin{equation}
C_{PRF}(\qorig,D) = C_{QE}(\qorig,D) + C_R(\qfinal,D),
\label{eq:cprf}
\end{equation}
where the cost metric $C_R$ is based on previous work showing that the sum of the lengths of the posting lists that need to be accessed for each query is strongly correlated with query response times \cite{moffat_pipelined_2007, macdonald_learning_2012}. 
Hence, for standard PRF, $C_{QE}(\qorig, D) = C_{R}(\qorig, D)$, and following \cite{moffat_pipelined_2007, macdonald_learning_2012,kulkarni_shard_2012, aly_taily_2013, kulkarni_selective_2015}, we define $C_R$ as follows.

\begin{definition}[Single-step retrieval cost]
The cost of retrieval for a given query $q$ is calculated as
\begin{equation}
C_{R}(\qorig,D) = \sum_{t \in \vec{q}} postings(t),
\end{equation}
where $postings(t)$ is the number of accessed postings in the inverted index for term $t$. 
\end{definition}

In the relevance model approach to PRF, the term selection step is constant for all PRF methods when we consider a fixed top-k number of documents. Hence, we discard this part of the cost because we are interested in relative costs.

\subsection{Expansion with External Corpus}
The use of an external corpus in query expansion has been studied in fields as far apart as blog search and Web search~\cite{arguello_document_2008, diaz_improving_2006, weerkamp_exploiting_2012}.
\citet{arguello_document_2008} addressed blog search with Wikipedia as an alternative query expansion corpus with significant improvements.
Freebase has been used for feedback to offer a wide coverage for past events and entities~\cite{fan_improving_2015}. 
In light of our goal, we aim to use \textit{the most up-to-date, reliable, and concise external corpus}.

To create a microblog expansion corpus there are several possible strategies. Sampling posts from Twitter accounts picked at random can lead to a low quality expansion collection.
Instead, using multiple authoritative news sources lends redundancy to the system since the same news story is often reported by multiple news sources.
Hence, since many news outlets use Twitter for the dissemination of news articles, we propose to listen the stream of news headlines directly from their Twitter profile pages (i.e., \emph{timelines}).
This corpus can be several orders of magnitude smaller than the target retrieval corpus. 

The implemented approach relies on an external news corpus covering multiple authoritative sources for expanding the query (we used 70 news sources). 
The news corpus is highly dynamic and is maintained up-to-date as new documents are arriving to be indexed -- current events are reported live as they unfold by online news sources. 
As a consequence of this dynamic environment, the \textit{query expansion corpus age and time span} will play a major role in the quality of the expansion corpus.

The use of news sampled from Twitter covers the information seeking behavior of users in the microblog search scenario. This is nicely tied to the natural topical bias of each query, suggesting that partitioning the expansion corpus into news verticals will bring greater benefits in terms of precision and expansion cost.

\subsection{\prfv{}: Pseudo-Relevant Vertical Feedback}

The federated query expansion architecture, stems from a new understanding of how temporal and topical information is searched in microblogs.
To take full advantage of the external expansion corpus the organization of documents into index shards is fundamental.
This architectural decision influences the latency and efficiency of the query expansion process.
Uniform sharding distributes the work across all machines so that it can be done more quickly, but it does not reduce the total work done and all the shards are involved for every search query.
Previous research found that topic-based shards offer the best balance of retrieval effectiveness and efficiency (query processing computation costs)~\cite{kulkarni_shard_2012, hafizoglu_efficiency_2017}.
Hence, we organize news sources into topic-based verticals, see Table \ref{tab:verticals}.

Queries are routed through a \emph{broker} to a subset of the most useful verticals. 
To make this decision, the broker keeps a \textit{central sample index} (CSI) of all verticals to select the few verticals to search for each query.
This reduces the amount of work done for each query and since the selected verticals can be searched in parallel this approach can be much faster.
In a \emph{cooperative}~\cite{shokouhi_federated_2011} federated search environment, global corpus statistics can be accessed by each federated search engine and by the \emph{broker} and therefore merging results from multiple verticals is straightforward.

Pseudo-Relevant Vertical Feedback (PRVF) is a query expansion architecture that uses an external corpus organized into verticals to efficiently select expansion terms. 
In the proposed approach,
the query expansion corpus is organized into a set of verticals 
$\IndexVerticalSet = \set{\indexverticalseq}$
from which a resource selection method
selects the most likely set $sel(\qorig)$,
which are then searched in parallel.
Formally, we wish to compute 
\begin{equation}
\selected{\origQuery} = \set{\indexvertical_{m(1)} \indexvertical_{m(2)} \cdots \indexvertical_{m(\numselected{\origQuery})}},
\label{eq:selq}
\end{equation}
where
$m: \set{1 \cdots \numselected{\origQuery}} \to \{1 \cdots \indexverticalseqlength\}$
is a mapping function
that indicates the set of $\numselected{\origQuery}$ verticals selected
given $\origQuery$,
which are the most promising,
in terms of relevance,
from the full set of $\indexverticalseqlength$ verticals.

To select the verticals, a resource selection algorithm either (i) uses a \emph{centralized sample index} (CSI) which indexes a representation sample $\corpus_{CSI}$~\cite{shokouhi_central-rank-based_2007, kulkarni_shard_2012} of each vertical's documents, or (ii) uses the term statistics~\cite{aly_taily_2013} of each vertical index.
With CSI based algorithms, we retrieve $R_k(\qorig, \corpus_{CSI})$, the top-$k$ documents from collection $\corpus_{CSI}$ in response to the initial query $\qorig$, using the \emph{query-likelihood} retrieval model. The verticals with more results in this sample are then selected for the feedback retrieval step. The key details of the implemented resource selection algorithms CRCS~\cite{shokouhi_central-rank-based_2007}, Rank-S~\cite{kulkarni_shard_2012} and Taily~\cite{aly_taily_2013} are in Section 2.

Finally, the top-$k$ documents retrieved from the verticals selected $sel(\qorig)$ in response to $\qorig$ are merged and used for feedback, \emph{vertical feedback}, to build the expansion language model $\qexp$,
\begin{equation}
R_k(\origQuery,\IndexVerticalSet) = \bigcup_{i=1}^{\numselected{\origQuery}} R_k(\origQuery, 
\indexvertical_{m(i)})
\end{equation}
which is interpolated with the original query model $\qorig$. This set of documents is then used to expand the original query, thus, ending the computation of $\qfinal$.

\subsubsection{Computational cost of PRVF}
It is worth recalling equation~\ref{eq:cprf}, where we defined the cost of standard PRF. Now, with Pseudo-Relevant Vertical Feedback (\prfv{}), the query expansion cost $C_{QE}$ is associated to $C_{VF}$, the cost of performing \textit{vertical feedback}. Formally, the PRVF cost is defined as
\begin{align}
\cprfv(q) &= C_{VF}(\qorig, \corpus_V) + C_R(\qfinal,\corpus),    
\label{eq:cprfv}
\end{align}
where $C_{VF}$ is the cost of expanding $\qorig$ on a vertical architecture and $C_R$ is the computational cost for searching the full index with the final query. 
The $C_{VF}$ efficiency measure proposed in~\citet{aly_taily_2013} accounts for two separate costs:
\begin{equation}
C_{VF}(\qorig, \corpus_V) = C_{SEL}(\qorig, \corpus_{V}) + C_{VR}(\qorig,\corpus_V)
\end{equation}
where $C_{SEL}(\qorig, \corpus_{CSI})$ is the cost of the resource selection algorithm, and $C_{VR}(\qorig,\corpus_V)$ (defined later) is the cost of retrieving documents in parallel from the selected verticals $sel(\qorig)$.
The cost of resource selection $C_{SEL}(\qorig)$ depends on the type of the resource selection algorithm used:
\begin{equation}
\cselection{\origQuery} =
\begin{cases} 
\hfill CSI(\origQuery)    \hfill & \text{ if sample-document} \\
\hfill \VerticalSet       \hfill & \text{ if vocabulary-based} \\
\end{cases}
\end{equation}
where $\VerticalSet$ is the total number of verticals
and $CSI(\origQuery) = \cretrieval{\origQuery}{\corpus_{CSI})}$
the number of postings accessed in the CSI
for all the query terms in $\origQuery$
considering a sample-document resource selection algorithm.
In the vocabulary-based resource selection algorithms~\cite{aly_taily_2013}, typically since a single look-up operation is performed, it is set to the total number of verticals $C_{SEL}(\qorig) = |\corpus_V|$.

\begin{definition}[Parallel retrieval cost]
In a vertical search scenario, $C_R$ is calculated for a given query $\qorig$ using the number of postings that the vertical search has to access as follows:
\begin{equation}
\begin{split}
C_{VR}(\qorig,\corpus_V) &= \sum_{i=1}^{|sel(\qorig)|} C_R(\qorig, \corpus_{V_{m(i)}})\\
&= \sum_{i=1}^{|sel(\qorig)|} \sum_{t \in \vec{\qorig}} postings_{\corpus_{V_{m(i)}}}(t)
\end{split}
\label{eq:crdv}
\end{equation}
where $sel(\qorig)$ are the verticals selected by a resource selection algorithm and $C_R(\qorig, \corpus_{V_{m(i)}})$ is the number of accessed postings in vertical $i$ for all the terms in the initial query $\qorig$.
\end{definition}

\subsubsection{Query response latency}
A federated search architecture also affords faster response times
via parallel work since multiple verticals can be searched in parallel.
Considering the set of documents $D$, the latency metric $C_{Lat}$ employed by \citet{kulkarni_selective_2015}, quantifies the longest execution path $C_{L}$ for a given query $\qorig$, assuming a distributed query processing framework,
\begin{equation}
\begin{split}
C_{Lat}(\qorig,\corpus_V) &= C_{SEL}(\qorig, \corpus_{V}) + C_L(\qorig,\corpus_V)\\
					  &= C_{SEL}(\qorig, \corpus_{V}) + \max_{i=1}^{|N|} \sum_{t \in \vec{\qorig}} postings_{\corpus_{V_{m(i)}}}(t)
\end{split}
\label{eq:clat}
\end{equation}
where $postings_{\corpus_{V_{m(i)}}}(t)$ is the number of accessed postings in the inverted index for term $t$ in vertical $i$, for a total of $N$ verticals in a federated search environment.

\subsection{Costs Comparison}
\label{sec:costs}
The computational cost of the \prfv{} approach is strongly correlated to the cost of retrieving candidate documents in a federated search retrieval system.
There might be non-negligible differences in the cost of the queries generated using different corpora for expansion.
That said, if the number of terms in the expanded query is fixed for all methods, the cost of retrieving the final results $C_R(\qfinal)$ can be assumed to be of the same order of magnitude for all methods presented.
Therefore, the first part of the cost equations \Cref{eq:cprf} and \Cref{eq:cprfv}, i.e., the cost of the query expansion process, $C_{QE}$, can be used alone to compare both approaches in terms of average computational costs:
\begin{gather}
C_{VF}(\qorig, \corpus_{VF}) << C_R(\qorig,\corpus) \\
\sum_{i=1}^{|sel(\qorig)|} \sum_{t \in \vec{\qorig}} postings_{\corpus_{V_{m(i)}}}(t) << \sum_{t \in \vec{q}} postings(t)
\end{gather}

The cost of the initial retrieval when using the whole collection is much larger than the proposed alternatives.
Using an index built with the posts of news sources provides a high-quality coverage of microblog user interests. Further computational gains are obtained by organizing news sources into topical verticals.
The hypothesis is that a $\prfv{}$ offers the lowest query expansion computational cost, and can provide comparable retrieval effectiveness to standard PRF techniques that use the whole corpus for feedback. Experiments will now examine this hypothesis.

\begin{table}[!t]
\caption{Verticals and sources.}
\centering
\begin{threeparttable}
\begin{tabularx}{0.9\linewidth}{ r X }
\toprule
Vertical ($\corpus_{V_i}$) & Twitter accounts                                                                                              \\ \midrule
general & abc, ap, bbcnews, bbcworld, cbsnews, cnn, cnni, foxnews, huffingtonpost, latimes, nprnews, nytimes, reuters, reutersuk, usatoday, mashable \\
politics & huffpostpol, politico, theeconomist, washingtonpost, wsj \\
technology & arstechnica, cnet, gizmodo, techcrunch, wired, wireduk, thenextweb, techrepublic, cnet, gigaom, macworld \\
sports & bbcsport, sinow, eurosport, eurosportuktv, sportscenter, espn \\
music & clash\_music, rollingstone, nme, spinmagazine, stereogum, billboard, altpress, pitchfork \\
movies & americancine, thr, nytmovies, bbcfilms, totalfilm, guardianfilm, backstage, empiremagazine, filmcomment, timeoutfilm, sightsoundmag \\
entertainment & time, ew, variety, vanityfair, uncutmagazine \\
science & livescience, popsci, wiredscience, nasa, natgeo, newscientist \\
breaking & bbcbreaking, breakingnews, cnnbrk \\
\bottomrule
\end{tabularx}
\footnotesize
\begin{tablenotes}
	\item \prfv(taily) selected 2.19 verticals on average for both TREC 2013 and TREC 2014 query sets. \prfv(ranks) selected 2.22 and 1.81 verticals on average for the TREC 2013 and TREC 2014 queries, respectively.
\end{tablenotes}
\end{threeparttable}
\label{tab:verticals}
\end{table}

\section{Experimental Methodology}

\subsection{Datasets}

\subsubsection{Microblog datasets. }
We use the Tweets2013 corpus and the topics from the TREC 2013 and TREC 2014 Microblog track~\cite{lin_overview_2013}.
Tweets2013 is a microblog posts collection (approx. 240 million \emph{tweets}) created by listening via Twitter's streaming API over the period: 1~February, 2013 -- 31~March, 2013 (inclusive).
NIST provided relevance judgments on a three-point scale of \enquote{informativeness}: not relevant, relevant and highly relevant.

\subsubsection{Vertical Expansion Corpus.}
Informed by previous work~\citep{rosa_topical_2011} and the categories presented in Twitter's sign up process, we created the following verticals: \emph{general, politics, entertainment, technology, breaking, movies, science, sports, music} and assigned each source to a vertical as shown in \Cref{tab:verticals}.
To build this corpus we selected 70 accounts from reliable news sources on Twitter.
We selected accounts from publications that are reputable and that are also popular on Twitter (i.e. have a high number of followers).

We obtained the set of posts published\footnote{\url{http://datasets.novasearch.org/tweets2013-newssources/}} by crawling the timelines of these news sources on the period covered by the Tweets2013 corpus: 1~February, 2013 -- 31~March, 2013 (inclusive), therefore we named it $NewsSources$ (140,087 documents).
A smaller collection (16,687 documents) is used as CSI for the evaluation of the resource selection algorithms and \prfv{}.
The collection was crawled using Twitter's $\sim$1\% and $\sim$10\% samples simultaneously. Documents from both streams were added to the index, which resulted in a volume of documents that is about 12\% of the size of $NewsSources$.

\subsection{Retrieval Methods}
To compare PRVF to previous research we considered 1 method without query expansion, 2 methods with expansion on the main corpus, and 5 methods with external expansion corpus. Performance was assessed in terms of MAP, NDCG@30, recall and computational cost (retrieval cost and latency).

\noindent\textbf{\noprf} is a \emph{query-likelihood} retrieval model with Dirichlet smoothing ($\mu = 2500$).

\noindent\textbf{PRF} is a standard pseudo-relevance feedback method that uses the whole search index for feedback.
The RM3 pseudo-relevance feedback algorithm~\cite{lavrenko_relevance_2001} is used  for all the methods based on PRF because it was shown to be very effective in previous microblog retrieval research and it has similar information requirements and computational characteristics to other PRF algorithms.
In all the PRF based methods, for feedback, we use the top 50 documents retrieved for each query $q$ using the \emph{query-likelihood} retrieval model, language modeling with Dirichlet smoothing ($\mu = 2500$).
For PRF.wiki the number of documents used for feedback was reduced to 10 articles.
The top-retrieved documents are then used to generate an expanded query $\qexp$ with a length of 20 terms.
The expansion terms $\qexp$ are interpolated with the original query terms $\qorig$ with equal weight ($\lambda = 0.5$).

\noindent\textbf{CLRM} (Condensed List Relevance Models) is an approach, based on relevance models, recently proposed by \citet{diaz_condensed_2015} that essentially re-ranks the list of results retrieved by the initial query using the expanded query generated with the same list.

\noindent\textbf{PRF.wiki} is a pseudo-relevance feedback baseline that uses an external index of the English Wikipedia article pages. It was used for expansion in blog search by \citet{arguello_document_2008} with significant improvements in effectiveness.
We process a Wikipedia dump\footnote{enwiki-20130102-pages-articles.xml}, which dates from just before the microblog evaluation dataset, using \emph{wikiextractor}\footnote{\url{https://github.com/attardi/wikiextractor}} to obtain clear indexable text.

\noindent\textbf{\prfs} is a pseudo-relevance feedback baseline that uses the whole $NewsSources$ dataset as an external expansion corpus.

\noindent\textbf{\prfv(taily)} can adjust the number of selected verticals $|sel(\qorig)|$ dynamically for each query.
It was parameterized with the values ($n = 400$ and $v = 50$) recommended by \citet{aly_taily_2013}.

\noindent\textbf{\prfv(ranks)} also adjusts the number of selected verticals $|sel(\qorig)|$ dynamically for each query.
To limit the number of verticals similarly for Rank-S we set the threshold $minRanks = 1e^{-6}$ .

\noindent\textbf{\prfv(crcs)} inspects a fixed number of verticals: $|sel(\qorig)| = \{1,2,3\}$.
For instance, \prfv(crcs1) corresponds to expanding the queries using only the top vertical selected by CRCS.

\section{Results and Discussion}
The evaluation is organized as follows: we start by analyzing the efficiency (\Cref{sec:efficiency}), biases (\Cref{sec:bias}) and effectiveness (\Cref{sec:effectiveness}) of \prfv{} and \prfs{} methods. 
Lastly, we compare the standard implementation of PRF based on re-retrieval with the recently proposed implementation Condensed List Relevance Models (CLRM) \cite{diaz_condensed_2015} based on re-ranking (\Cref{sec:clrm}).

\subsection{Cost Analysis of PRF Methods}
\label{sec:efficiency}
\begin{figure}[t]
	{\centering 
		\subcaptionbox{TREC 2013.\label{fig:mapcres-trec2013}}{%
			\includegraphics[trim={7 5 8 10},clip,width=0.23\textwidth]{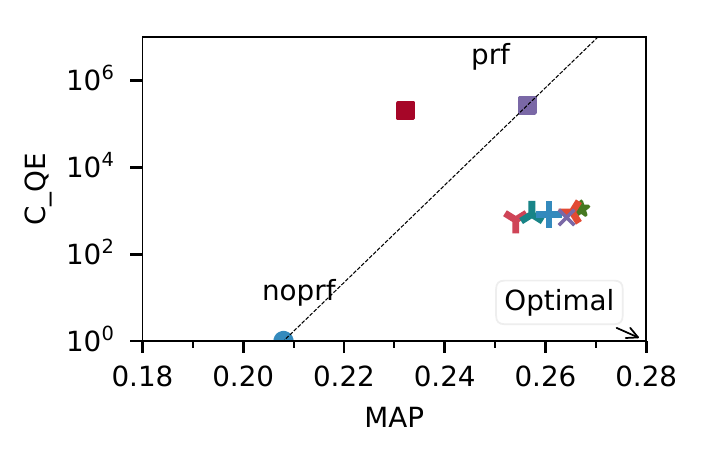}%
		}
		\subcaptionbox{TREC 2014.\label{fig:mapcres-trec2014}}{%
			\includegraphics[trim={7 5 8 10},clip,width=0.23\textwidth]{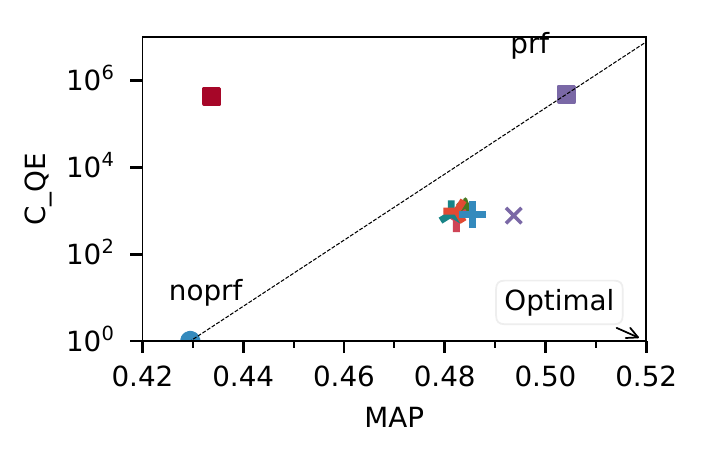}%
		}\\
		\subcaptionbox{TREC 2013.\label{fig:ndcg30cres-trec2013}}{%
			\includegraphics[trim={7 5 8 10},clip,width=0.23\textwidth]{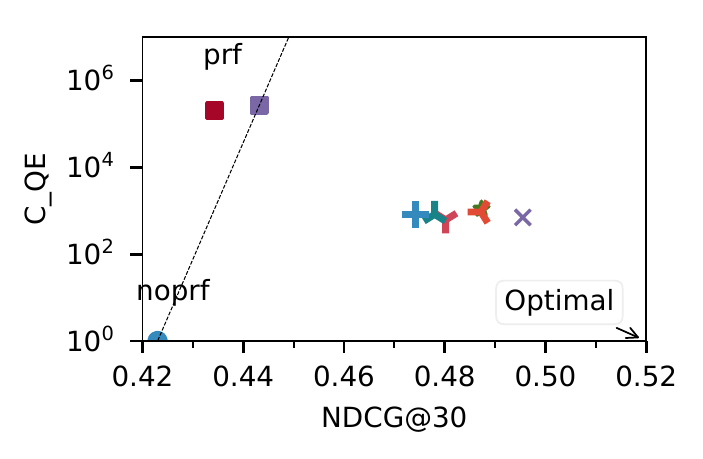}%
		}
		\subcaptionbox{TREC 2014.\label{fig:ndcg30cres-trec2014}}{%
			\includegraphics[trim={7 5 8 10},clip,width=0.23\textwidth]{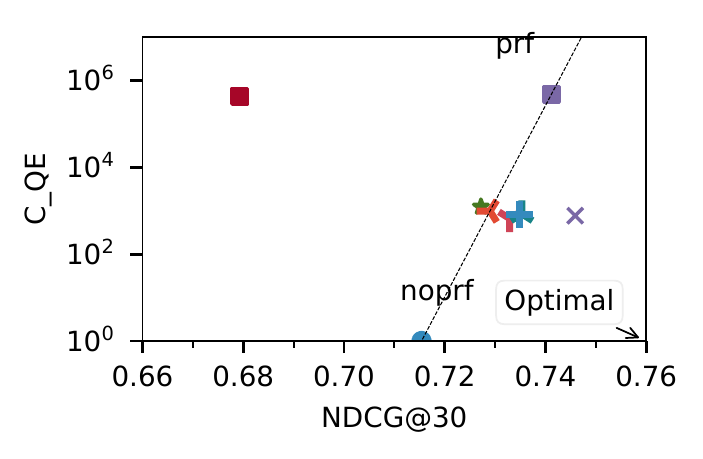}%
		}
		\\
		\includegraphics[trim={0 0 0 0},clip,width=0.6\columnwidth]{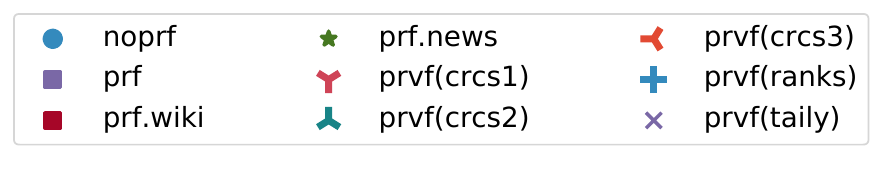}\par}
	\caption{Comparison of retrieval cost ($C_{QE}$) versus retrieval effectiveness (using precision metrics MAP and NDCG@30).}
	\label{fig:xy_plots}
\end{figure}

In this paper we focus on computational cost reduction for the initial retrieval, necessary in PRF-based query expansion. In this section we measure the computational cost, $C_{QE}$, as the number of accessed posting lists for query expansion. See \Cref{sec:costs} for the query expansion costs of each method.
\Cref{fig:xy_plots} shows the trade-offs between the cost, $C_{QE}$, and the corresponding results on retrieval metrics on the TREC Microblog datasets.
$C_{QE}$ is represented in the y-axis in log-scale to get an overview of how PRVF compares to the \noprf{} and standard PRF baselines.
The x-axis is represents either the MAP or the NDCG@30 retrieval metric.
Since the objective is to lower $C_{QE}$ and get better retrieval precision, the desired method would fall below the dashed line that goes from \noprf{} to PRF, towards the bottom right corner.

The \prfv{} methods have always a lower computational cost $C_{QE}$ than \prfs, clustering just below it in the graphs.
Even though the cost of the \prfv{} methods is considerably lower, the MAP results obtained are near those of \prfs.
Therefore, \prfs{} can be a good predictor for the expected retrieval effectiveness for the \prfv{} architecture (see \Cref{fig:mapcres-trec2014}).

While the initial expectation is that more computational cost should translate into a better ranking, some \prfv{} methods provide a better top rank with a lower computational cost.
The NDCG@30 results obtained with \prfv{} methods are similar or better than PRF (see \Cref{fig:ndcg30cres-trec2014}).
\prfv(taily) outperformed the other methods in NDCG@30 in the TREC 2014 queries.
However, on the TREC 2013 queries \prfv(crcs3) outperformed the \prfv(taily) method slightly as it can be seen in \Cref{fig:ndcg30cres-trec2013}.

All the proposed approaches were three orders of magnitude less computationally expensive than the PRF baseline.
The \prfv{}-based methods are the most efficient since for each query they search only the verticals that are most promising.
We found that query expansion response times can be halved on both datasets compared to the \prfs{} baseline as measured by $C_{Lat}$, which takes into account the parallelism afforded by the distributed architecture.

\subsection{Quality of Expansion Corpus}
\label{sec:bias}
In this section we analyze potential biases in the vertical expansion corpus and the importance of the expansion corpus age and time span.
Since \prfv{} uses documents from news sources for query expansion we might improve the chances of retrieving tweets from these sources.
To make sure that bias is not improving the results unfairly we counted the number of documents marked as relevant in the main index (\textit{TREC 2013 and 2014}) that are in the expansion corpus (\textit{NewsSources}).
The overlap was only 9 relevant documents in TREC 2013 and 13 relevant documents for TREC 2014.
Thus, we did not find any evidence that the choice of news sources affords any kind of unfair advantage. 

\begin{figure}[t]
	{\centering 
		\subcaptionbox{TREC 2013.\label{fig:delay-trec2013}}{%
			\includegraphics[trim={7 5 8 7},clip,width=0.23\textwidth]{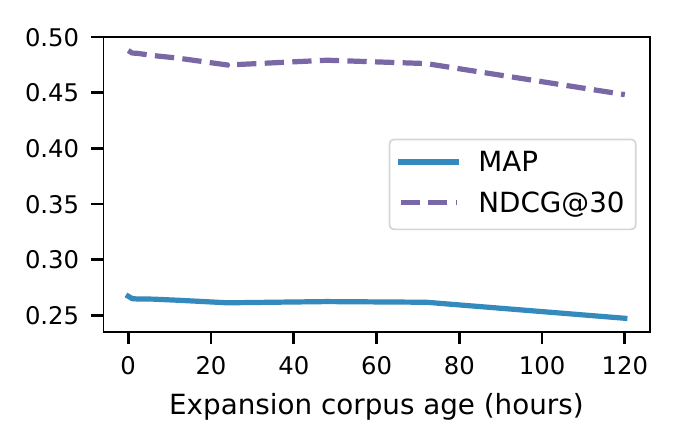}%
		}
		\subcaptionbox{TREC 2014.\label{fig:delay-trec2014}}{%
			\includegraphics[trim={7 5 8 7},clip,width=0.23\textwidth]{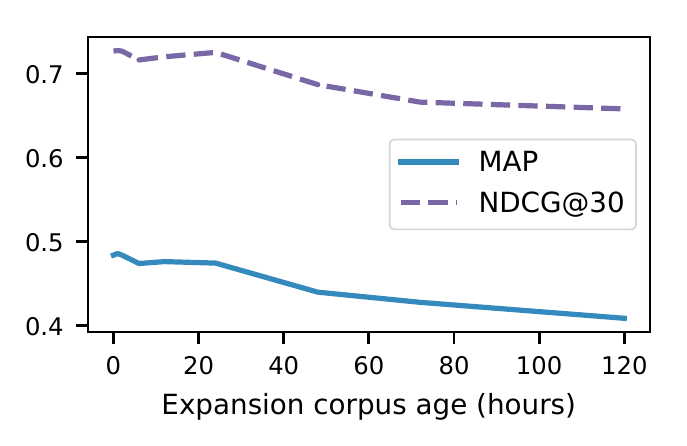}%
		}
		\\
		\subcaptionbox{TREC 2013.\label{fig:timespan-trec2013}}{%
			\includegraphics[trim={7 5 8 7},clip,width=0.23\textwidth]{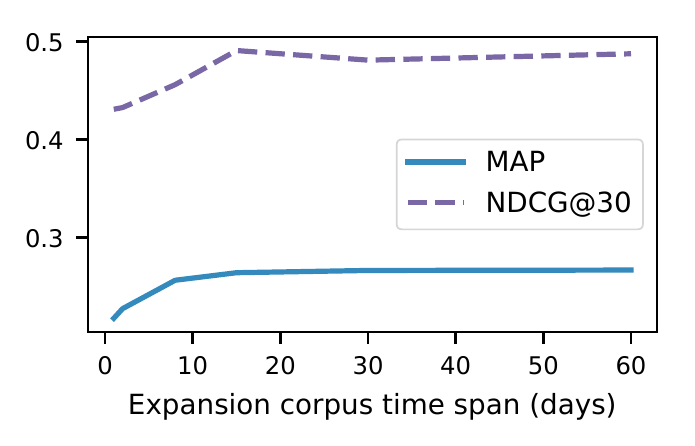}%
		}
		\subcaptionbox{TREC 2014.\label{fig:timespan-trec2014}}{%
			\includegraphics[trim={7 5 8 7},clip,width=0.23\textwidth]{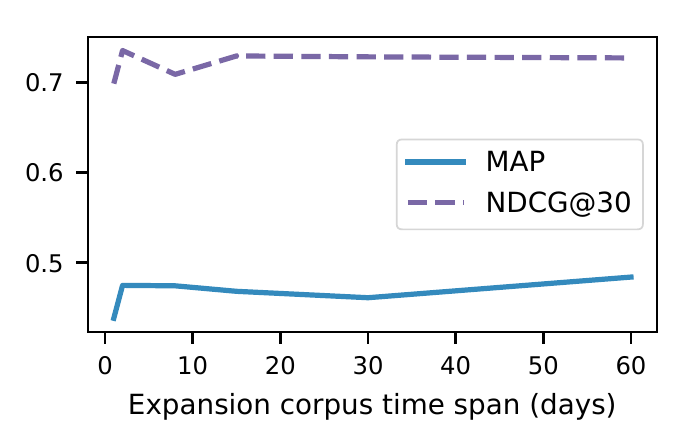}%
		}
	}
	\vspace{-1.07\baselineskip}
	\caption{Analysis of expansion corpus age and time span.}
	\label{fig:corpus_age}
\end{figure}

A key aspect of the PRVF architecture is its ability to cope with multiple information streams that are constantly feeding the query expansion corpus. In \Cref{fig:delay-trec2013} and \Cref{fig:delay-trec2014} we observe how the \textit{expansion corpus age}, i.e. the difference between queries timestamp and the most recent document timestamp, is clearly linked to the decay in retrieval precision.
The time span of the expansion corpus is also examined in \Cref{fig:timespan-trec2013} and \ref{fig:timespan-trec2014} -- here we can observe that it might be sufficient to keep only the last 15 days for query expansion.
This confirms the initial assumption that in microblog search it is critical to use an up-to-date expansion corpus. In addition, we found that it does not seem to be necessary to keep an expansion corpus with a long time span to answer most queries effectively.

\subsection{Retrieval Effectiveness of PRF Methods}
\label{sec:effectiveness}
We show the detailed results of our evaluation on \Cref{tab:exp-results-2013} and \Cref{tab:exp-results-2014}.
We present the average over all queries for the ranking metrics MAP and NDCG@30.
The average computational cost of each approach is presented in the $C_{QE}$ column.
After the $C_{QE}$, in parentheses, we show cost reduction in relation to \prfs.
The language model baseline, \noprf, does not use pseudo-relevance feedback therefore $C_{QE}$ does not apply.

A federated query expansion architecture also affords faster response times via parallel work.
When multiple verticals are searched in parallel, the query expansion process waits for all the response of all verticals to proceed with the term selection phrase.
Therefore we measure response time using $C_{Lat}$ which can give us the maximum amount of work done by any vertical searched in parallel.
We find that query expansion times are halved in both datasets compared to the \prfs{} baseline.

The PRF baseline improved the retrieval effectiveness metrics considerably over \noprf{} and PRF.wiki.
In the TREC 2013 queries the PRF baseline improved on MAP over \noprf{} by 23.3\% and NDCG@30 by 4.8\%, and the MAP improvement was statistically significant.
For TREC 2014 queries PRF improved on MAP by 17.4\% and NDGC@30 by 3.6\% over \noprf{}, and the MAP improvement was statistically significant.
However, the average cost of the query expansion process using standard PRF for TREC 2013 and TREC 2014 was very high at 269k and 488k postings, respectively.

We followed onto expansion methods with external corpus.
Using a recent Wikipedia corpus for feedback, PRF.wiki, only provided an improvement of MAP on the TREC 2013 queries.
However, for the TREC 2014 queries, NDCG@30 was 5\% lower than the \noprf{} baseline.
In addition, even though the Wikipedia corpus is smaller than the search corpus the average computational costs were still very high at 202k and 434k accessed postings for TREC 2013 and TREC 2014, respectively.
We conclude then that using Wikipedia for query expansion in microblogs can harm retrieval effectiveness when the expansion collection is not up-to-date.

Using the \textit{NewsSources} corpus for feedback, PRF.news, provided a significant improvement in terms of retrieval precision and computational cost. However, to fully verify the aforementioned hypothesis, we examined the impact of creating expansion verticals. 

\begin{table}[t]
\caption{TREC 2013 dataset results.}
\centering
\begin{threeparttable}
\begin{tabularx}{1\columnwidth}{p{2cm} *{4}{Y}}
\toprule
			& $C_{Lat}$  & $C_{QE}$   & MAP       & NDCG    \\
\midrule
\multicolumn{5}{l}{w/o External Expansion Corpus} 
\\
\midrule
\noprf         
&  0
&  0
& 0.2080   & 0.4230
\\
PRF            
&  269175
&  269175
& 0.2564   & 0.4432
\\
\midrule
\multicolumn{5}{l}{w/ External Expansion Corpus} 
\\
\midrule
PRF.wiki      
&  201892
&  201892
& 0.2323   & 0.4343
\\
\prfs          
&  1110
&  1110
& \bftab0.2671\bc& 0.4873\bc
\\
\midrule
\multicolumn{5}{l}{w/ External Vertical Expansion Corpus} 
\\
\midrule
\prfv(crcs1)               
&  631 \scriptsize{-43.2$\%$}
&  631 \scriptsize{-43.2$\%$}
& 0.2541\bc   & 0.4802\bc
\\
\prfv(crcs2)               
&  653 \scriptsize{-41.2$\%$}
&  845 \scriptsize{-23.9$\%$}
& 0.2573\bc   & 0.4780\bc
\\
\prfv(crcs3)               
&  655 \scriptsize{-41.0$\%$}
&  956 \scriptsize{-13.9$\%$}
& 0.2653\bc   & 0.4872\bc
\\
\prfv(ranks)              
&  673 \scriptsize{-39.4$\%$}
&  903 \scriptsize{-18.6$\%$}
& 0.2607\bc   & 0.4742\wc
\\
\prfv(taily)               
&  \bftab509 \scriptsize{-54.1$\%$}
&  \bftab703 \scriptsize{-36.7$\%$}
& 0.2642\bc  & \bftab0.4955\bc
\\
\bottomrule
\end{tabularx}
	\footnotesize
	\begin{tablenotes}
		\item \emph{Symbols \textnormal{\wc} and \textnormal{\bc} stand for a statistically non-inferior result to PRF with $p < 0.05$ and $p < 0.01$ respectively, according to a non-inferiority test \citep{walker_understanding_2011}.}
	\end{tablenotes}
\end{threeparttable}

\label{tab:exp-results-2013}
\end{table}

In the group of methods that use an external vertical expansion corpus, the \prfv(taily) method was the most balanced in the TREC 2013 queries with a $C_{QE}$ of only 703, which corresponds to a cost reduction of 36.7$\%$ over \prfs, around 2.2$\times$ faster.
It had one of the highest MAP results (3.0\% higher than PRF) and improved 27.0\% over \noprf{} (statistically significant).
It also had the second best NDCG@30 result, improving 1.7\% over \prfs{} and 11.8\% over the standard PRF approach.

\prfv(taily) provided the best balance for the TREC 2014 queries as well.
It obtained a 14.7\% improvement in MAP over \noprf{} (statistically significant) and a 4.4\% improvement in NDCG@30 with a computational cost of only $C_{QE} = 773$.
\prfv(taily) was around 2.2$\times$ faster than searching the whole news index \prfs{} a cost reduction of 37.6$\%$.
The highest MAP results for TREC 2013 were obtained with \prfv(crcs3).
Because a fixed number of verticals are searched for each query (3), which is higher than \prfv(taily)'s average, the cost reduction is smaller (13.9$\%$ over \prfs).
Last, but not least, \prfv(taily) was the fastest method, delivering the lowest latency ($C_{Lat}$ column), halving the latency of PRF.news.

\begin{table}[t]
	\caption{TREC 2014 dataset results.
	}
	\centering
	\begin{threeparttable}
		\begin{tabularx}{1\columnwidth}{p{2cm} *{4}{Y} *{2}{Z}}
			\toprule
			& $C_{Lat}$  & $C_{QE}$   & MAP       & NDCG    \\
			\midrule
			\multicolumn{5}{l}{w/o External Expansion Corpus} 
			\\
			\midrule
			\noprf         
			&  0
			&  0
			& 0.4295   & 0.7154
			\\
			PRF            
			&  487547
			&  487547
			&\bftab 0.5042   & 0.7412
			\\
			\midrule
			\multicolumn{5}{l}{w/ External Expansion Corpus} 
			\\
			\midrule
			PRF.wiki      
			&  434233
			&  434233
			& 0.4338   & 0.6793
			\\
			\prfs          
			&  1239
			&  1239
			& 0.4841   & 0.7272\wc
			\\
			\midrule
			\multicolumn{5}{l}{w/ External Vertical Expansion Corpus} 
			\\
			\midrule
			\prfv(crcs1)               
			&  661 \scriptsize{-46.7$\%$}
			&  661 \scriptsize{-46.7$\%$}
			& 0.4823   & 0.7329\bc
			\\
			\prfv(crcs2)               
			&  734 \scriptsize{-40.8$\%$}
			&  848 \scriptsize{-31.6$\%$}
			& 0.4813   & 0.7353\bc
			\\
			\prfv(crcs3)               
			&  734 \scriptsize{-40.8$\%$}
			&  982 \scriptsize{-20.7$\%$}
			& 0.4824   & 0.7290\wc
			\\
			\prfv(ranks)              
			&  734 \scriptsize{-40.8$\%$}
			&  821 \scriptsize{-33.7$\%$}
			& 0.4856   & 0.7348\bc
			\\
			\prfv(taily)               
			&  \bftab575 \scriptsize{-53.6$\%$}
			&  \bftab773 \scriptsize{-37.6$\%$}
			& 0.4927\bc   & \bftab0.7470\bc
			\\
			\bottomrule
		\end{tabularx}
		\footnotesize
		\begin{tablenotes}
			\item \emph{Symbols \textnormal{\wc} and \textnormal{\bc} stand for a statistically non-inferior result to PRF with $p < 0.05$ and $p < 0.01$ respectively, according to a non-inferiority test \citep{walker_understanding_2011}.}
		\end{tablenotes}
	\end{threeparttable}
	\label{tab:exp-results-2014}
\end{table}

\begin{table}[ht]
\caption{Retrieval results using CLRM on microblog datasets.}
\centering
\begin{tabularx}{1\columnwidth}{p{2cm} *{6}{Y}}
\toprule
                    & \multicolumn{3}{c}{TREC 2013}        & \multicolumn{3}{c}{TREC 2014} \\
\cmidrule(l){2-4}
\cmidrule(lr){5-7}
                         & MAP    & NDCG   & Recall   & MAP       & NDCG  & Recall   \\
\midrule
\noprf                   & 0.2080 & 0.4230 & 0.5188 & 0.4295 & 0.7154 & 0.6994 \\
PRF                      & 0.2564 & 0.4432 & 0.5764 & \bftab0.5042 & 0.7412 & \bftab0.7860 \\
CLRM          			 & 0.2276 & 0.4423  & 0.5188 & 0.4718 & 0.7416& 0.6994 \\
\midrule
PRF.wiki     			 & 0.2323 & 0.4343  & 0.5689   & 0.4338      & 0.6793   & 0.7443 \\
\prfv(taily)             & \bftab0.2642 & \bftab0.4955  &  \bftab0.5921 & 0.4927 & \bftab0.7470&  0.7818 \\
\bottomrule
\end{tabularx}
\label{tab:clrm-results}
\vspace{-5mm}
\end{table}

\subsection{Re-ranking PRF and Short Text Documents}
\label{sec:clrm}

In \Cref{tab:clrm-results} we present retrieval effectiveness metrics for CLRM and other PRF implementations based on re-retrieval.
We also present the \textit{recall} metric on the top 1000 results (number of relevants @ 1000/total of relevants).
We found that CLRM always outperformed No-PRF in MAP and NDCG@30. 
However, since CLRM does not perform a re-retrieval it re-ranks the documents retrieved by the initial query, therefore its recall  is equal to the query likelihood baseline \noprf{}.
Due to this, CLRM is not able to achieve the same level of retrieval effectiveness of the implementations based on re-retrieval.
Even though the generated expanded query is the same for CLRM and PRF, PRF was more effective.

In short-text document indexes some relevant documents that are ranked at the top by a re-retrieval implementation might be missing from the initial retrieval using the original query terms only.
In addition, some relevant documents might contain only a few of the original query terms, a problem that is exacerbated by the short size of the documents in a microblog corpus.
Therefore, in short text datasets an implementation of pseudo-relevance feedback based on re-retrieval might be preferred to achieve similar retrieval effectiveness to standard PRF.

The PRF.wiki method, based on re-retrieval, was able to retrieve more relevant documents than CLRM with a higher recall in both datasets.
However, this higher recall did not translate into better results since the query expansions generated from Wikipedia were less effective for ranking.

\prfv(taily) approach generates query expansions using a more efficient federated query expansion architecture over an external news corpus.
The quality of the expansions generated by this method can be attested from its high recall and better precision than the standard PRF approach.

\section{Conclusion}
\label{sec:discussion}
In this paper, we studied an efficient method for pseudo-relevance feedback that organizes large collections of documents, published by a set of news sources into news verticals.
The evaluation of the proposed architecture led us to the following concluding points.

\vspace{0.5mm}
\noindent
\textbf{Federated QE.} PRVF architecture is an efficient federated QE architecture for microblog search, where the expansion corpus is live and has new documents arriving from news sources in a streaming fashion.
This approach outperformed the retrieval effectiveness of using the non-partitioned news index (\prfs) and PRF.

\vspace{0.5mm}
\noindent
\textbf{Low-cost and effective PRF.} The best balance between efficiency and effectiveness was obtained using \prfv(taily), which was relatively more robust than other approaches while using on average fewer verticals.
\prfv(taily) achieved the highest results in effectiveness metrics for both the TREC 2013 and TREC 2014 query sets. \prfv(crcs3) was the best in terms of MAP on TREC 2013, but at a slightly higher computational cost.
This indicates that resource selection algorithms that can dynamically limit the number of verticals searched are more suitable for this task.

\vspace{0.5mm}
\noindent\textbf{Quality of the query expansion corpus.} Understanding the properties of the search domain is key to ensure the quality of the expansion corpus. In microblog search, news sources as the ones in \cref{tab:verticals} guarantee a good and up-to-date coverage of user interests. This is a critical aspect that is addressed by domain-specific knowledge.

\vspace*{-2.5mm}
\begin{acks}
This work has been partially funded by the \grantsponsor{CMUP}{CMU Portugal}{} research project GoLocal Ref. \grantnum{CMUP}{CMUP-ERI/TIC/0033/2014}, by the \grantsponsor{EU H2020}{H2020 ICT}{} project COGNITUS with the grant agreement n\textsuperscript{o} \grantnum{EU H2020}{687605}, and by the \grantsponsor{FCT}{FCT}{} project NOVA LINCS Ref. \grantnum{FCT}{UID/CEC/04516/2013}.
\end{acks}

\vspace*{-2.0mm}
\newcommand{\shownote}[1]{\unskip}
\newcommand{\showDOI}[1]{\unskip}
\newcommand{\showURL}[1]{\unskip}


%
%

\end{document}